\documentclass[twocolumn]{jetc20132e}
\usepackage{cite}
\usepackage{bm}  
\usepackage[utf8]{inputenc}
\usepackage{textcomp}
\usepackage[T1]{fontenc} 
\usepackage{mathrsfs}
\usepackage{amsmath}

\title{Long-range interacting systems and the Gibbs-Duhem equation
}

\author{Ivan Latella and Agust\'in P\'erez-Madrid 
    \affiliation{
Departament de F\'{i}sica Fonamental, Facultat de F\'{i}sica, Universitat de Barcelona\\ 
Mart\'{i} i Franqu\`{e}s 1, 08028 Barcelona, Spain\\
E-mail: ilatella@ffn.ub.edu}
}%

\date {
{\small The generalized Gibbs-Duhem equation is obtained for systems with long-range interactions in $d$ spatial dimensions.
We consider that particles in the system interact through a slowly decaying pair potential of the form $1/r^\nu$ with $0\leq\nu\leq d$.
The local equation of state is obtained by computing the local entropy per particle and using the condition of local thermodynamic equilibrium. This local equation of state turns out to be that of an ideal gas. Integrating the relation satisfied by local thermodynamic variables over the volume, the equation involving global magnitudes is derived. Thus, the Euler relation is found and we show that it is modified by the addition of a term proportional to the total potential energy. This term is responsible for the modification of the Gibbs-Duhem equation. We also point out a close relationship between the thermodynamics of long-range interacting systems and the thermodynamics of small systems introduced by Hill.
}
}

\begin{document}

\newcommand{\vect}[1]{\bm{#1}}
\newcommand{\nomen}[1]{\makebox[10mm][l]{#1}}
\newcommand{\e}{\mathrm{e}}
\newcommand{\dif}{\mathrm{d}}
\newcommand{\iunit}{\mathrm{i}}

\maketitle

\section*{INTRODUCTION}

Long-range interacting systems have received considerable attention in recent years due to their remarkable dynamical and statistical behavior \cite{Campa:2009,Bouchet:2010}.
Self-gravitating systems \cite{Antonov:1962,Lynden-Bell:1968,Thirring:1970,Padmanabhan:1990,Chavanis:2002:a}, two-dimensional vortices \cite{Chavanis:2002:c}, nuclear physics \cite{Chomaz:2002} and also toy mo\-dels such as the Hamiltonian mean field model \cite{Dauxois:2002} are examples of systems presenting such a behavior.
These systems are intrinsically non-additive and may have negative heat capacity in the micro\-canonical ensemble leading to ensemble in\-equivalence \cite{Ruffo:2002,Bouchet:2005,Campa:2009}.
   
Systems with long-range interactions are characterized by slowly-decaying pair potentials through which the constituent parts of the system interact at large distances.
When particles interact all with the same coupling, the absence of screening causes inhomogeneous configurations (except for the limiting case of non-decaying interactions), in equilibrium or meta-equilibrium, where the formalism of thermodynamics can be applied.
Unlike the case of short-range interacting systems, where many features are well understood, there is a lack of complete knowledge about the dynamical and statistical properties of systems with long-range interactions.

To be more specific, a definition of what we mean by long-range potentials can be formulated as follows: a potential that at large distances decays as $1/r^\nu$ is formally said to be long-range if $\nu\leq d$, where $d$ is the dimension of the embedding space \cite{Campa:2009,Bouchet:2010}.
Below we will give an argument that justifies why this particular range for the power of the decaying potential deserves special attention.
Thus, here we are concerned with the study of the thermodynamics of $d$-dimensional systems with these power-law interaction potentials ($0\leq\nu\leq d$) by using the mean field approximation.

In the mean field approach it is implicitly assumed that the number of particles is large enough so that the system can be treated as a continuous medium and the description of the relevant physical quantities is assumed to depend on the density (or distribution function) in the one-particle phase space.
Despite the fact that correlations are ignored in the mean field approach, this model offers a mathematical tool for a suitable treatment of self-interactions in the system.
It turns out to be very accurate in the thermodynamic limit, except near the critical points where the system undergoes a phase transition or collapses \cite{Chomaz:2002,deVega:2002:a,deVega:2002:b,Ispolatov:2001:a,Ispolatov:2001:b}.
Although the validity of the mean field solutions strongly depends on the control parameters used to specify the thermodynamic state of the system, the functional form of any thermodynamic potential in the mean field limit is the same in each ensemble representation.

Quite remarkably, de Vega and S\'anchez obtained the local equation of state of the self-gravitating gas assuming local hydrostatic equilibrium \cite{deVega:2002:b}. This equation coincides with that of an ideal gas.
Additionally, the same local equation of state is found using the condition of local hydrostatic equilibrium for a system with arbitrary long-range interactions in the mean field limit \cite{Chavanis:2011}.

We will see that the local ideal gas equation of state of the system can be obtained by computing the local entropy per particle and using the condition of local thermodynamic equilibrium \cite{Latella:2013}.
Although the result is the same, this procedure is conceptually different.
The local entropy per particle (and also other thermodynamic potentials) can be obtained from the volume of the phase space using the saddle point approximation or, equivalently, using the one-particle distribution function approach. After volume integration of the relation satisfied by local quantities, an equation involving global thermodynamic quantities can be found. This is the Euler relation, which for systems with the interaction potentials considered here, is mo\-dified by the addition of an extra term containing the total potential energy. This reflects the fact that an extra degree of freedom, proportional to the potential energy, has to be considered to formulate a thermodynamic description of systems with long-range interactions. The formal structure of thermodynamic relations for systems with long-range interactions is thus the same as the corresponding one for small systems. It can be seen that the total potential energy plays the role of the subdivision potential introduced by Hill \cite{Hill:
1963} to treat small systems. Because systems with long-range interactions may also be considered as small, this connection can be seen as more than a formal mapping between mathematical relations. 

\section*{LOCAL EQUATION OF STATE}

Consider a $d$-dimensional system of $N$ point-like particles of mass $m$ enclosed in a spherical container of volume $V$ (and radius $R$), which interact through a pair interaction potential that at large distances behaves as
\begin{equation}
\phi_{ij}=\kappa |\vect{q}_i-\vect{q}_j|^{-\nu}\, ,
\label{interaction:potential}
\end{equation}
where $\kappa$ is a coupling constant, $\vect{q}_i$ is the coordinate of particle $i$, $i=1,2,\dots, N$, and $0\leq\nu\leq d$.
In $d=1$ the container is a $0$-sphere which is the pair of end-points of the line segment of length $2R$, in $d=2$ the $1$-sphere is formed by the points at the boundary of a circumference of radius $R$, and so on.
The Hamiltonian is given by $H_N=E_0+W$, where $E_0$ is the kinetic energy and $W=\sum_{i>j}^N\phi_{ij}$ is the total potential energy.

In the microcanonical description, the state of the system is characterized by a fixed value of the total energy $E$ and the number of microstates in full $2d$-dimensional phase space is given by $\Sigma(E)=\left(2\pi\hbar\right)^{-dN}(N!)^{-1}\int_{E>H_N}\dif \tau$, where $\hbar$ is the reduced Planck's constant and $\dif \tau$ is the volume element in phase space.
Thus, the microcanonical entropy reads $S(E)=k_\mathrm{B}\ln\Sigma(E)$, where $k_\mathrm{B}$ is Boltzmann's constant.
To compute the entropy in the mean field limit, the volume of the system is divided in cells and, after integrating over momentum, the configurational integrals in $\Sigma(E)$ become summations over all possible occupation number distributions.
In the limit $N\rightarrow\infty$, the discrete occupation number distributions become continuous fields and summations become a functional integration over the number density $n(\vect{x})$, where now $\vect{x}$ represents the spatial components of a single point in the one-particle configuration space.
This functional integration is solved with the saddle-point approximation in such a way that the number density that maximizes the entropy (hence defining the equilibrium configurations) is given by \cite{Latella:2013}
\begin{equation}
n(\vect{x})=\lambda_T^{-d}\exp\left[\frac{\mu-\Phi(\vect{x})}{k_\mathrm{B}T}\right] ,
\label{number:density:micro}
\end{equation}
where $\lambda_T=\left[2\pi\hbar^2/(mk_\mathrm{B}T)\right]^{1/2}$ is the thermal wavelength, $\mu$ is the chemical potential and $T$ is the temperature.
Here we have introduced the self-consistent potential $\Phi(\vect{x})$ which takes the form
\begin{equation}
\Phi(\vect{x})=\int n(\vect{x}')\phi(\vect{x},\vect{x}')\,\dif^d\vect{x}',
\label{potential}
\end{equation}
where now $\phi(\vect{x},\vect{x}')=\kappa |\vect{x}-\vect{x}'|^{-\nu}$ describes the interaction between particles in the one-particle configuration space. Notice that $\Phi(\vect{x})$ depends explicitly on $\nu$ and, therefore, so does the density.
As a result, the microcanonical mean field entropy becomes \cite{Latella:2013}
\begin{equation}
S= k_\mathrm{B}\int n(\vect{x})\left[-\ln\left(n(\vect{x})\lambda_T^d\right)+\frac{2+d}{2}\right]\, \dif^d\vect{x} .
\label{microcanonical:S}
\end{equation}
In addition, in terms of $\Phi(\vect{x})$ the total potential energy takes the usual form
\begin{equation}
W=\frac{1}{2}\int n(\vect{x})\Phi(\vect{x})\,\dif^d\vect{x} ,
\label{potential:energy:2}
\end{equation}
and the total energy reads
\begin{equation}
E=\frac{dk_\mathrm{B}T}{2}\int n(\vect{x})\,\dif^d\vect{x}+\frac{1}{2}\int n(\vect{x})\Phi(\vect{x})\,\dif^d\vect{x}. 
\end{equation}

The details of the above calculations can be found in \cite{Latella:2013}, where the used method is based on a previous work  \cite{deVega:2002:a,deVega:2002:b} concerning self-gravitating systems ($\nu=1$).
This method and the validity of the expressions for the thermodynamic quantities that are obtained in the mean field approach rest on the assumption that the interactions are long-ranged.
In other words, it is assumed that the main contribution to the interaction energy of a particle is due to distant particles rather than to its immediate neighbors.
To see that this is fulfilled with $0\leq\nu\leq d$, we now come back to the argument that justifies the formal definition of long-range potentials. We will follow \cite{Campa:2009}.
Let us consider a particle placed at the origin of the $(d-1)$-dimensional sphere of radius $R$.
In order to estimate the energy $\epsilon$ of the particle due to the interaction with the rest of particles in the bulk, let us assume that particles are homogeneously distributed so that $n(\vect{x})=$ constant. We also assume that there is a short-distance cutoff $\delta\ll R$ describing the scale where short-range interactions have to be considered.
Writing $r=|\vect{x}-\vect{x}'|$ we have
\begin{equation}
\epsilon =\int_\delta^R\dif^dr\,n\,\frac{\kappa}{r^\nu}=\left\{
\begin{array}{ll}
n\kappa\Omega_{(d)}{\displaystyle\frac{R^{d-\nu}-\delta^{d-\nu}}{d-\nu}},&\quad \nu\neq d \\
n\kappa\Omega_{(d)}\ln\left(R/\delta\right),&\quad \nu= d\, 
\end{array}
\right. ,
\end{equation}
where $\Omega_{(d)}=2\pi^{d/2}/\Gamma(d/2)$ is the solid angle factor, $\Gamma(x)$ being the Gamma function.
On the one hand, we see that if $0\leq \nu\leq d$, the integral is dominated by the contribution coming from its upper limit and then by long-range interactions.
The energy $\epsilon$ grows as $\epsilon\propto V^\sigma$ if $0\leq \nu<d$ (logarithmically in the marginal case $\nu=d$), where $\sigma\equiv 1-\nu/d$ is the long-range parameter, and hence the total energy scales as $E\propto V^{\sigma+1}$. On the other hand, if $\nu>d$ the energy $\epsilon$ remains finite for $\delta/R\ll 1$ and consequently $E\propto V$.
This is the usual scaling of extensive systems where interactions are short-ranged ($\nu>d$).
Therefore, the mean field approach can suitably describe interactions in the system only if they are long-ranged and the formulation considered here is not valid for short-range potentials.
It is worth noting that, in general, the equilibrium or meta\-equilibrium configurations have to be obtained by using numerical calculations due to the non-trivial functional relation between the potential and the density.
For the case of self-gravity, the potential satisfies the Poisson-Boltzmann equation. Therefore, thermodynamic quantities can be expressed in closed form in terms of its solution.
However, this equation has to be solved numerically as well.

Local thermodynamic quantities can be defined taking into account our ability to write the entropy and the energy as integrals over the volume.
In view of (\ref{microcanonical:S}), the local entropy per particle is given by
$s(\vect{x})=k_\mathrm{B}\left[-\ln\left(n(\vect{x})\lambda_T^d\right)+\frac{2+d}{2}\right]$.
This is a Sackur-Tetrode-type entropy per particle written in terms of local variables.
In the same way, the local kinetic energy and local energy per particle take the form $e_0=\frac{d}{2} k_\mathrm{B}T$ and $e(\vect{x})=e_0+\frac{1}{2}\Phi(\vect{x})$, respectively.
Thus we have
\begin{align}
&S= \int n(\vect{x})s(\vect{x})\, \dif^d\vect{x} ,\\
&E= \int n(\vect{x})e(\vect{x})\, \dif^d\vect{x} ,\\
&E_0= \int n(\vect{x})e_0\, \dif^d\vect{x} .
\end{align}
We also introduce the local volume per particle defined by $v(\vect{x})=1/n(\vect{x})$, so that the local entropy can be written as a function of the local kinetic energy and this local volume, $s=s(e_0,v)$.
From the condition of local thermodynamic equilibrium we have \cite{deGroot:1962,Glansdorff:1977},
\begin{equation}
\frac{1}{T}=\left(\frac{\partial s}{\partial e_0}\right)_v\qquad\mbox{and}\qquad\frac{p}{T}=\left(\frac{\partial s}{\partial v}\right)_{e_0}\, , 
\label{local:equilibrium}
\end{equation}
which leads to the local equation of state \cite{Latella:2013}
\begin{equation}\label{local:equation:state}
p(\vect{x})=n(\vect{x})k_\mathrm{B}T. 
\end{equation}
This implies that particles at a certain point $\vect{x}$ behave as an ideal gas, but the pressure and density vary from point to point (except in the limiting case of non-decaying interaction potential, i.e. $\sigma=1$).
Since local interactions (short-ranged) are neglected in comparison to the interaction with distant particles, this result is physically consistent.
Therefore, the system behaves locally as a free gas under the action of an external field created by the particles in the bulk.
The same result can be found by considering local hydrostatic equilibrium \cite{deVega:2002:a,deVega:2002:b,Chavanis:2011}.
This is achieved by equating the gradient of pressure to the force density:
\begin{equation}
\nabla  p(\vect{x})=n(\vect{x})\nabla\Phi(\vect{x}).
\end{equation}

In addition to the functional form of the local variables, we can also write down an explicit relation among them.
Taking into account the expression for the local entropy and introducing the ideal gas chemical potential
$\mu_0(\vect{x})=k_\mathrm{B}T\ln\left[n(\vect{x})\lambda_T^d\right]$, one deduces
\begin{equation}\label{local:relation:1}
Ts(\vect{x})=e_0+p(\vect{x})v(\vect{x})-\mu_0(\vect{x})\, .
\end{equation}
The same equation could have been deduced if as a starting point one assumes that the system locally behaves as an ideal gas.
Since $\mu=\mu_0(\vect{x})+\Phi(\vect{x})$, equation (\ref{local:relation:1}) can also be rewritten in the form
\begin{equation}
Ts(\vect{x})=e(\vect{x})+p(\vect{x})v(\vect{x})-\mu+\frac{1}{2}\Phi(\vect{x})\, .
\label{local:relation}
\end{equation}
Notice the presence of the last term on the r.h.s. in the above expression; it is due to the definition of the local energy $e(\vect{x})$. As we will see in the next section, such a term will give rise to an extra term in the Euler relation and therefore, in the Gibbs-Duhem equation.

An alternative way to address the local description of the system is by considering the distribution function $f(\vect{x}, \vect{p})$ defined in one-particle phase space, where $\vect{p}$ is the momentum of a particle. The distribution function is normalized so that
$N=\int f(\vect{x},\vect{p})\, \dif^d\vect{x}\, \dif^d\vect{p}$. In order to obtain equilibrium or meta\-equilibrium configurations, using variational calculus one looks for the distribution function which maximizes the Boltzmann entropy
$S=-k_\mathrm{B}\int f(\vect{x},\vect{p}) \ln(f(\vect{x},\vect{p})/f_{\mathrm{c}}) \, \dif^d\vect{x}\, \dif^d\vect{p}$, where $f_{\mathrm{c}}$ is a constant fixing the origin of the entropy. Consequently, the distribution function that maximizes the entropy (at least locally) turns out to be the Maxwell-Boltzmann distribution with the self-consistent potential (\ref{potential}), and the number density is given by (\ref{number:density:micro}). Besides, the constant $f_{\mathrm{c}}$ is chosen so that the entropy coincides with the one in the micro\-canonical description: $f_{\mathrm{c}}=\mathrm{e}/(2\pi\hbar)^d$.

Thus, any local magnitude per particle $q(\vect{x})$ associated with the global quantity $Q$ can be defined according to
\begin{equation}\label{quantity}
n(\vect{x})q(\vect{x})\equiv\int f(\vect{x},\vect{p}) \mathcal{Q}(\vect{x},\vect{p})\, \dif^d\vect{p}
\end{equation}
provided
$Q=\int f(\vect{x},\vect{p}) \mathcal{Q}(\vect{x},\vect{p}) \, \dif^d\vect{x}\, \dif^d\vect{p}$.
For instance, in the case of the Boltzmann entropy one takes
$\mathcal{Q}(\vect{x},\vect{p})=-k_\mathrm{B}\ln(f(\vect{x},\vect{p})/f_{\mathrm{c}})$. After integrating over momentum using the Maxwell-Boltzmann distribution, the resulting local entropy per particle $s(\vect{x})$ is the same as the one found in the microcanonical ensemble using the mean field approximation. Therefore, the condition of local thermodynamic equilibrium leads to the same local equation of state \cite{Latella:2013}, equation (\ref{local:equation:state}).

\section*{GLOBAL THERMODYNAMIC RELATIONS}

Once the relation between thermodynamic variables is esta\-blished at a local level, the corresponding relation between global variables is obtained by integrating over the volume.
Concretely, we want to obtain the corresponding Euler relation from which a generalization of the Gibbs-Duhem equation can be deduced.
In this way, multiplying both sides of (\ref{local:relation}) by $n(\vect{x})$ and integrating over the volume yields
\begin{equation}
TS=E+\frac{2}{d}E_0-\mu N +W\, ,
\label{global:relation:1} 
\end{equation}
where we have used that $\int p(\vect{x})\, \dif^d\vect{x}=k_\mathrm{B}T\int n(\vect{x})\, \dif^d\vect{x}=\frac{2}{d}E_0$.
We now introduce the pressure evaluated at the boundary of the $d$-dimensional system, $P\equiv p(\vect{x})|_{\vect{x}\; \in\;\mathrm{boundary}}$.
To proceed further, the pressure $P$ has to be related to (\ref{global:relation:1}).
This can be done if the global equation of state is taken into account, which can be computed by rescaling the energy in the micro\-canonical density of states and using the usual thermodynamic relations for the total entropy \cite{deVega:2002:a,deVega:2002:b,Latella:2013}.
The rescaled energy reads $\Lambda\equiv ER^\nu/\left(|\kappa|N^2\right)$ \cite{deVega:2002:a,Ispolatov:2001:b}.
As a result, the global equation of state takes the form
\begin{equation}
\frac{PV}{Nk_\mathrm{B}T}= 1+\nu\frac{W}{dNk_\mathrm{B}T}.
\label{equation:state}
\end{equation}
Therefore, with the help of (\ref{equation:state}), equation (\ref{global:relation:1}) can be rewritten in  such a way that
\begin{equation}
TS=E+PV-\mu N +\sigma W,
\label{global:relation:2} 
\end{equation}
which is the Euler relation for the systems discussed here.
Moreover, by differentiating (\ref{global:relation:2}) one gets
\begin{equation}
T\dif S=\dif E+P\dif V -\mu\dif N +\sigma\dif W -N\dif \mu-S\dif T+V\dif P
\label{differential_relation} 
\end{equation}
and since
$T\dif S = \dif E+P\dif V -\mu\dif N$
one obtains \cite{Latella:2013}
\begin{equation}
\sigma\dif W= S\dif T - V\dif P + N\dif \mu\, ,
\label{Gibbs-Duhem}
\end{equation}
which is the generalized Gibbs-Duhem equation for long-range interacting systems.

The marginal case $\nu=d$ corresponds to systems with long-range parameter $\sigma=0$. In such a case, the Euler relation and the generalized Gibbs-Duhem equation reduce to the usual one in the thermodynamics of short-range interactions. For $\sigma\neq0$, the temperature, chemical potential and pressure are independent variables, so that $W=W(T,P,\mu)$. The thermodynamic relations in terms of partial derivatives of $W$ have been considered in \cite{Latella:2013} and verified for the case of a self-gravitating gas ($\sigma=2/3$) and for a system with spatially uniform interactions ($\sigma=1$).

In what follows we will see that the thermodynamic relations satisfied by systems with long-range interactions can be mapped to the corresponding relations satisfied by small systems introduced by Hill \cite{Hill:1963}. Hill's small systems bear this name because they are composed by a small (non-macroscopic) number of particles. The systems we have considered above are macroscopic in the sense that the number of particles is assumed to be infinite. However, the latter are small in the sense that the range of the interactions is large compared to the size of the system. Due to this finiteness, an extra degree of freedom has to be considered to account for a complete thermodynamic description. This extra degree of freedom in long-range systems is $\sigma W$ while for systems with small number of particles it is incorporated through the subdivision potential $\mathscr{E}$. 
 
In the formalism introduced by Hill \cite{Hill:1963}, an ensemble of non-interacting small systems is considered. Thus, the subdivision potential accounts for the energy gained by the system when the number of members of the ensemble varies. For any single small system one has
\begin{align}
&TS=E+P V-\mu N- \mathscr{E}\, ,\label{Hill:1}\\
&T\dif S=\dif E+P\dif V-\mu\dif N\, ,\label{Hill:2}\\
&\dif \mathscr{E}=-S\dif T+V\dif P-N\dif \mu\, .\label{Hill:3}
\end{align}
As it can be seen, these relations are exactly the same as those we have obtained for systems with long-range interactions if the identification $\mathscr{E}=-\sigma W$ is made.

To conclude, it is important to stress here, as was mentioned by Hill, that the use of different environmental variables, i.e. control parameters, would lead to different descriptions of the thermodynamic phenomena when small systems are considered \cite{Hill:1963}. The same occurs in long-range interacting systems when different ensemble representations are not equivalent. We have considered only the microcanonical ensemble, but the results apply also for the canonical and grand canonical ensembles. This is because in the mean field approximation the thermodynamic potentials all have the same functional form in the different ensembles \cite{Latella:2013}. However, the critical points where the mean field approximation ceases to be valid are characteristic of each ensemble. 

\section*{DISCUSSION}

By integration of the relation among the different local thermodynamic variables over the volume of the system, we find the corresponding equation satisfied by the global variables.
It is shown that the potential energy enters as a thermodynamic variable which modifies the global thermodynamic equations.
That is, the Euler relation is modified if the system possesses long-range interactions and takes the form $TS=E+PV-\mu N +\sigma W$.
As a result, we find a generalized Gibbs-Duhem equation which relates the potential energy to the intensive variables:
$\sigma\dif W= S\dif T - V\dif P + N\dif \mu$.
For the marginal case where the power of the decaying interaction potential is equal to the dimension of the embedding space, the usual Gibbs-Duhem equation is recovered.
Therefore, when long-range interactions are present in the system, the intensive variables become independent due to the freedom introduced by the potential energy. The potential energy naturally depends on the intensive variables. 
We also emphasize that this deviation from standard thermodynamics is similar to what happens with Hill's thermodynamics of small systems. 

\begin{acknowledgment}
This work was supported by the Spanish Government under grant FIS2011-22603.
I L acknowledges financial support through an FPI scholarship (BES-2012-054782) from the Spa\-nish Government.
\end{acknowledgment}

\begin{nomenclature}
\entry{\nomen{$d$}}{Dimension of the embedding space}%
\entry{\nomen{$\dif\tau$}}{Volume element in full phase space}%
\entry{\nomen{$e(\vect{x})$}}{Local energy per particle at point $\vect{x}$}%
\entry{\nomen{$E$}}{Total energy}%
\entry{\nomen{$\mathscr{E}$}}{Subdivision potential}%
\entry{\nomen{$e_0$}}{Local kinetic energy per particle}%
\entry{\nomen{$E_0$}}{Total kinetic energy}%
\entry{\nomen{$f(\vect{x}, \vect{p})$}}{Distribution function in one-particle phase space}%
\entry{\nomen{$f_{\mathrm{c}}$}}{Constant in Boltzmann entropy $=\mathrm{e}/(2\pi\hbar)^d$}%
\entry{\nomen{$\hbar$}}{Reduced Planck's constant = $1.054\,571\,726(47)\times10^{-34}\,\mathrm{J}\,\mathrm{s}$}%
\entry{\nomen{$H_N$}}{$N$-particle Hamiltonian}%
\entry{\nomen{$k_\mathrm{B}$}}{Boltzmann's constant = $1.380\,648\,8(13)\times 10^{-23}\,\mathrm{J}\,\mathrm{K}^{-1}$}%
\entry{\nomen{$m$}}{Mass of a particle}%
\entry{\nomen{$n(\vect{x})$}}{Number density at point $\vect{x}$}%
\entry{\nomen{$N$}}{Number of particles}%
\entry{\nomen{$\vect{p}$}}{Momentum in one-particle phase space}%
\entry{\nomen{$p(\vect{x})$}}{Local pressure at point $\vect{x}$}%
\entry{\nomen{$P$}}{Pressure at the boundary of the system}%
\entry{\nomen{$q(\vect{x})$}}{Generic local magnitude per particle}%
\entry{\nomen{$Q$}}{Generic global magnitude}%
\entry{\nomen{$\mathcal{Q}(\vect{x}, \vect{p})$}}{Generic magnitude in one-particle phase space}%
\entry{\nomen{$\vect{q}_i$}}{Coordinate of particle $i$}%
\entry{\nomen{$r$}}{Interparticle distance}
\entry{\nomen{$R$}}{Radius of the spherical container}%
\entry{\nomen{$s(\vect{x})$}}{Local entropy per particle at point $\vect{x}$}%
\entry{\nomen{$S$}}{Total entropy}%
\entry{\nomen{$T$}}{Temperature}%
\entry{\nomen{$v(\vect{x})$}}{Local volume per particle at point $\vect{x}$}%
\entry{\nomen{$V$}}{Volume of the system}%
\entry{\nomen{$W$}}{Total potential energy}%
\entry{\nomen{$\vect{x}$}}{Position in one-particle configuration space}%
\entry{\nomen{$\delta$}}{Short-distance cutoff}%
\entry{\nomen{$\Gamma(x)$}}{Gamma function of $x$}%
\entry{\nomen{$\epsilon$}}{Energy of a particle at the center of the system}%
\entry{\nomen{$\kappa$}}{Generic coupling constant}%
\entry{\nomen{$\Lambda$}}{Rescaled energy}%
\entry{\nomen{$\lambda_T$}}{Thermal wavelength}%
\entry{\nomen{$\mu$}}{Chemical potential}%
\entry{\nomen{$\mu_0(\vect{x})$}}{Ideal gas chemical potential at point $\vect{x}$}%
\entry{\nomen{$\nu$}}{Power of the decaying pair interaction potential}%
\entry{\nomen{$\sigma$}}{Long-range parameter $=1-\nu/d$}%
\entry{\nomen{$\Sigma(E)$}}{Number of microstates}%
\entry{\nomen{$\phi(\vect{x},\vect{x}')$}}{Pair interaction potential between particles in one-particle phase space}%
\entry{\nomen{$\Phi(\vect{x})$}}{Potential at point $\vect{x}$}%
\entry{\nomen{$\phi_{ij}$}}{Pair interaction potential in full phase space}%
\entry{\nomen{$\Omega_{(d)}$}}{Solid angle factor for the $(d-1)$-sphere}%
\end{nomenclature}


\bibliographystyle{unsrt}
\bibliography{Student-paper-Latella-JETC2013}

\endmytext 

\end{document}